\documentclass[reprint,onecolumn,superscriptaddress,amsmath,amssymb,aps]{revtex4-2}
\usepackage[utf8]{inputenc}
\usepackage[T1]{fontenc}
\usepackage{graphicx}
\usepackage[x11names]{xcolor}
\usepackage{hyperref}
\hypersetup{ 
     colorlinks=true,
     urlcolor= Blue3,
     linkcolor= Blue3,
     citecolor=Green4,
}
\usepackage{cleveref}
\usepackage{booktabs}
\usepackage{tikz}
\usepackage{xparse}
\usepackage{notes2bib}
\usepackage[normalem]{ulem}

\NewDocumentCommand{\rot}{O{45} O{1em} m}{\makebox[#2][l]{\rotatebox{#1}{#3}}}
\crefname{equation}{Eq.}{Eqs.}
\Crefname{equation}{Equation}{Equations}
\crefname{figure}{Fig.}{Figs.}
\Crefname{figure}{Figure}{Figures}
\crefname{section}{Sec.}{Secs.}
\Crefname{section}{Section}{Sections}
\crefname{appendix}{Appendix}{Apps.}
\Crefname{appendix}{Appendix}{Apps.}
\crefname{paragraph}{Sec.}{Secs.}
\crefname{table}{Table}{Tables}

\newcommand{\smalldot}{\;\;\tikz\draw[black,fill=black] (0,0) circle (.5ex);}
\newcommand{\bigdot}{\;\;\tikz\draw[black,fill=black] (0,0) circle (1.ex);}

\begin{document}
\title{Quantum technologies for climate change: Preliminary assessment}

\author{Casey Berger}
\affiliation{Department of Physics, Smith College, Northampton, Massachusetts 01063, USA}
\affiliation{Boston University Rafik B. Hariri Institute for Computing and
Computational Science \& Engineering, Boston, Massachusetts 02215, USA}
\author{Agustin Di Paolo}
\affiliation{Département de physique, Institut quantique, Université de Sherbrooke, Sherbrooke, Québec, J1K 2R1, Canada}
\author{Tracey Forrest}
\affiliation{Transformative Quantum Technologies, University of Waterloo, Waterloo, Ontario, Canada}
\author{Stuart Hadfield}
\affiliation{USRA Research Institute for Advanced Computer Science, Mountain View, California 94043, USA}
\author{Nicolas Sawaya}
\affiliation{Intel Labs, Santa Clara, California 95054, USA}
\author{Michał Stęchły}
\affiliation{Zapata Computing Canada Inc., 325 Front St W, Toronto, ON, M5V 2Y1}
\author{Karl Thibault}
\affiliation{Département de physique, Institut quantique, Université de Sherbrooke, Sherbrooke, Québec, J1K 2R1, Canada}

\date{\today. Last updated \today.}

\begin{abstract}
    Climate change presents an existential threat to human societies and the Earth's ecosystems more generally. Mitigation strategies naturally require solving a wide range of challenging problems in science, engineering, and economics. In this context, rapidly developing \textit{quantum} technologies in computing, sensing, and communication could become useful tools to diagnose and help mitigate the effects of climate change. However, the intersection between climate and quantum sciences remains largely unexplored. This preliminary report aims to identify potential high-impact use-cases of quantum technologies for climate change with a focus on four main areas: simulating physical systems, combinatorial optimization, sensing, and energy efficiency. We hope this report provides a useful resource towards connecting the climate and quantum science communities, and to this end we identify relevant research questions and next steps.
\end{abstract}

\maketitle

\section*{Introduction}
\label{s: Introduction}
Emerging quantum technologies are anticipated to provide new approaches to challenging scientific and engineering problems~\cite{preskill_quantum_2018,alexeev2021quantum}. An important question is to what extent these technologies can provide advantages related to anthropogenic climate change in areas such as mitigation, adaptation and prediction~\cite{edenhofer2011renewable,ipcc2014_industry,turner2020climate,rolnick2019tackling,giani2021quantum}. Moreover, understanding the time scale in which quantum advantages could be realized is particularly important for pressing climate related challenges, and assessing the potential \textit{real-world} impact. Hence, identifying the most promising use cases for quantum technologies of the near future, weighted by potential impact, is a natural first step in this direction. This report seeks to expedite the formation of an interdisciplinary community of researchers toward elucidating these important questions in quantum climate science.

\textit{Q4Climate} is an initiative that brings together the research communities at the intersection of quantum and climate science, involving academia, government, and industry, with the aim of identifying how or where emerging quantum technologies may be applied to reduce the pace and impact of climate change. 

This report presents a preliminary high-level assessment of the potential of quantum technologies for tackling climate change, with a focus on impactful realizations in the near- to moderate-term. Our assessment is by no means exhaustive or complete, and intends to provide a basis for more thorough investigations of specific applications. More precisely, we identify and discuss four application areas: simulation of physical systems (Sec.~\ref{s: QSimulation}), combinatorial optimization (Sec.~\ref{s: QOptimization}), quantum sensing (Sec.~\ref{s: QSensing}), and (computational) energy efficiency (Sec.~\ref{s: EnergyEfficiency}). Our preliminary investigation mostly avoids applications with more demanding quantum resource requirements, for example, advanced quantum approaches for solving differential equations, or quantum approaches to machine learning; we reserve such topics for future iterations of this report. We emphasize that as quantum technology continues to develop, both in theory and in practice, the views of this document may change considerably. Indeed, we identify a number of relevant research questions and areas for further investigation (throughout and Sec.~\ref{s:conclusion}). Interested researchers are welcome and encouraged to get in touch with us at \href{www.q4climate.org}{www.q4climate.org}. 

\newpage
\section{Quantum Simulation \\ \scriptsize{Casey Berger, Agustin Di Paolo and Nicolas Sawaya}}
\label{s: QSimulation}

The advent of supercomputers brought massive advances in materials science and chemistry. However, and despite increasingly innovative algorithms and the growing accessibility of large-scale computing power, most materials science and chemistry problems remain intractable by standard or `classical' computers. For many of these problems, intractability is a mere consequence of their quantum-mechanical nature. Thus, quantum computing (QC) holds great promise for enabling advances in these areas \cite{cao2019quantum,bauer20_chemrev,mcardle20_revmodphys}.

Acknowledging the technological challenges that lie ahead, the goal of this chapter is (a) to highlight the technological and scientific subfields that would benefit from more powerful computation and (b) to categorize the types of simulations and algorithms relevant to climate and energy.

\subsubsection*{Opportunities ahead}
At the microscopic level of electrons and atoms, all physical behavior is governed by quantum mechanics. However, studying the quantum scale with 
classical computers faces major obstacles, as the time and memory required for exact simulation increases exponentially with the size of the system. For instance, in order to perform an exact simulation of $N$ quantum particles with $d$ states each, one generally needs to store and do linear-algebra operations on at least a $d^N$-dimensional space. For a relatively small problem of 100 quantum particles, this would require a memory size of at least $2^{100} \approx 10^{30}$, a quantity equivalent to $> 10^{89}$ terabytes that one could never hope to even store in a classical computer.

This intractable scaling could be avoided if one had access to a large enough quantum computer, where the amount of computing resources (number of qubits and processing time) can scale polynomially with~$N$. Thus, for large~$N$, the simulation possibilities offered by a quantum processor 
appear to be simply unmatched by their classical counterpart, something that could enable the solution of technologically important and difficult problems in physics and chemistry. 

However, the dream of quantum computers, 
assuming viable large-scale devices can be built and deployed over a reasonable timeframe, is not simply to speed up existing calculations. First, to a degree of accuracy that is impossible today, researchers could in principle simulate the behavior of, for example, new photovoltaics, battery materials and fuels \textit{before} spending massive time and investment in the laboratory. Though there are currently many useful materials and chemical simulations performed on traditional computers, their imprecise results can often be used only for qualitative conclusions or for eliminating poor candidates. Second, with a quantum computer one would be able to perform virtual screenings \cite{jain16_dft} of millions of molecules (or more) for  given applications, with high confidence in the accuracy of the results (as opposed to the mid or low confidence given by current supercomputer simulations). 

The question therefore appears not to be whether a quantum computer can aid in development of clean technology, but whether a robust enough quantum computer can be built in time. Despite the formidable challenges that lie ahead, in this chapter we outline ways in which a quantum computer could become a useful tool to help alleviate the climate crisis \textit{if} a viable machine is built in the next 10-20 years. 

\begin{table*}[t!]
\centering
\begin{tabular}{lccccccccccccc}
    & \rot{Solar energy \cite{Nayak19_solar} }
    & \;\;\rot{Wind energy \cite{thomas18_windblades,Matizamhuka18_windmagn} }
    & \;\;\rot{Batteries \cite{Dunn11_batteryrev} }
    & \;\;\rot{Industrial processes \cite{friedmann19_heat} }
    & \;\;\rot{Materials for the grid \cite{gur18_storageandgrid} }
    & \;\;\rot{Synthetic fuels (incl. electrolysis) \cite{hanggi19_syntheticfuel}  }
    & \;\;\rot{CO$_2$ capture \cite{Tcvetkov19_seques,BenMansour16_co2adsorp} }
    & \;\;\rot{Atmospheric science  \cite{seinfeld16_atmosbook} }
    & \;\;\rot{Nuclear power \cite{mit18_nuclear} }
    & \;\;\rot{Structural materials \cite{ipcc2014_industry} }
    & \;\;\rot{Biological enzymes  \cite{othman17_greenfuel} }
    & \;\;\rot{Plastics \& circular economy }
    & \;\;\rot{Agriculture }\\
    \midrule
    Gas-phase electronic structure \cite{szabo_book} & \;\;- & \;\;- & \;\;- & \;\;\tikz\draw[black,fill=black] (0,0) circle (1.ex); & \;\;- & \;\;\tikz\draw[black,fill=black] (0,0) circle (1.ex); & \;\;\tikz\draw[black,fill=black] (0,0) circle (1.ex); & \;\;\tikz\draw[black,fill=black] (0,0) circle (1.ex); & \;\;- & \;\;- & \;\;-  & \smalldot  & \bigdot \\
    Molecular dynamics \cite{markland18_md} & \;\;- & \;\;- & \;\;\tikz\draw[black,fill=black] (0,0) circle (.5ex);  & \;\;\tikz\draw[black,fill=black] (0,0) circle (1.ex); & \;\;- & \;\;\tikz\draw[black,fill=black] (0,0) circle (1.ex); & \;\;\tikz\draw[black,fill=black] (0,0) circle (1.ex); & \;\;\tikz\draw[black,fill=black] (0,0) circle (1.ex); & \;\;- & \;\;- & \;\;\tikz\draw[black,fill=black] (0,0) circle (1.ex); & \smalldot  & \smalldot \\
    Solution chemistry \cite{solvation_book}
& \;\;- & \;\;- & \;\;\tikz\draw[black,fill=black] (0,0) circle (1.ex); & \;\;- & \;\;- & \;\;\tikz\draw[black,fill=black] (0,0) circle (1.ex); & \;\;\tikz\draw[black,fill=black] (0,0) circle (1.ex); & \;\;\tikz\draw[black,fill=black] (0,0) circle (1.ex); & \;\;- & \;\;- & \;\;\tikz\draw[black,fill=black] (0,0) circle (1.ex); & \bigdot  & \bigdot \\
    Transition metal elements \cite{elfving20_indust} & \;\;\tikz\draw[black,fill=black] (0,0) circle (1.ex); & \;\;\tikz\draw[black,fill=black] (0,0) circle (1.ex); & \;\;\tikz\draw[black,fill=black] (0,0) circle (1.ex); & \;\;\tikz\draw[black,fill=black] (0,0) circle (1.ex); & \;\;\tikz\draw[black,fill=black] (0,0) circle (1.ex); & \;\;\tikz\draw[black,fill=black] (0,0) circle (1.ex); & \;\;- & \;\;- & \;\;\tikz\draw[black,fill=black] (0,0) circle (.5ex); & \;\;\tikz\draw[black,fill=black] (0,0) circle (.5ex); & \;\;\tikz\draw[black,fill=black] (0,0) circle (1.ex); & \bigdot  & \;\;- \\
    Lanthanides \& actinides \cite{Tatewaki17_relat} & \;\;\tikz\draw[black,fill=black] (0,0) circle (1.ex); & \;\;\tikz\draw[black,fill=black] (0,0) circle (1.ex); & \;\;\tikz\draw[black,fill=black] (0,0) circle (.5ex); & \;\;\tikz\draw[black,fill=black] (0,0) circle (.5ex); & \;\;\tikz\draw[black,fill=black] (0,0) circle (.5ex); & \;\;\tikz\draw[black,fill=black] (0,0) circle (.5ex); & \;\;- & \;\;- & \;\;\tikz\draw[black,fill=black] (0,0) circle (1.ex); & \;\;\tikz\draw[black,fill=black] (0,0) circle (.5ex); & \;\;- & \;\;-  & \;\;- \\
    Electronic band structure \cite{kaxiras08_book} & \;\;\tikz\draw[black,fill=black] (0,0) circle (1.ex); & \;\;- & \;\;\tikz\draw[black,fill=black] (0,0) circle (1.ex); & \;\;- & \;\;- & \;\;\tikz\draw[black,fill=black] (0,0) circle (1.ex); & \;\;- & \;\;- & \;\;- & \;\;- & \;\;- & \;\;-  & \;\;- \\
    Electron/hole diffusion constants \cite{kaxiras08_book} & \;\;\tikz\draw[black,fill=black] (0,0) circle (1.ex); & \;\;- & \;\;\tikz\draw[black,fill=black] (0,0) circle (1.ex); & \;\;- & \;\;- & \;\;\tikz\draw[black,fill=black] (0,0) circle (1.ex); & \;\;- & \;\;- & \;\;- & \;\;- & \;\;- & \;\;-  & \;\;- \\
    Vibrational and vibronic structure \cite{mcardle19_qvibr,sawaya21_spectr} & \;\;\tikz\draw[black,fill=black] (0,0) circle (.5ex); & \;\;- & \;\;\tikz\draw[black,fill=black] (0,0) circle (.5ex); & \;\;\tikz\draw[black,fill=black] (0,0) circle (.5ex); & \;\;- & \;\;\tikz\draw[black,fill=black] (0,0) circle (.5ex); & \;\;\tikz\draw[black,fill=black] (0,0) circle (.5ex); & \;\;\tikz\draw[black,fill=black] (0,0) circle (1.ex); & \;\;- & \;\;- & \;\;\tikz\draw[black,fill=black] (0,0) circle (1.ex); & \;\;-  & \bigdot \\
    Magnetism  \cite{Siouani19_mag} & \;\;\tikz\draw[black,fill=black] (0,0) circle (.5ex); & \;\;\tikz\draw[black,fill=black] (0,0) circle (1.ex); & \;\;- & \;\;\tikz\draw[black,fill=black] (0,0) circle (.5ex); & \;\;\tikz\draw[black,fill=black] (0,0) circle (1.ex); & \;\;- & \;\;- & \;\;- & \;\;\tikz\draw[black,fill=black] (0,0) circle (.5ex); & \;\;- & \;\;- & \;\;-  & \;\;- \\
    Nuclear structure and reactions \cite{Ebata20_nucstruc} & \;\;- & \;\;- & \;\;- & \;\;- & \;\;- & \;\;- & \;\;- & \;\;- & \;\;\tikz\draw[black,fill=black] (0,0) circle (1.ex); & \;\;- & \;\;- & \;\;-  & \;\;- \\
    Excited states of a Hamiltonian \cite{higgott19_vqd,jones19_discovspectra,motta19_qite,ollitrault19_eom} & \;\;\tikz\draw[black,fill=black] (0,0) circle (1.ex); & \;\;- & \;\;- & \;\;\tikz\draw[black,fill=black] (0,0) circle (.5ex); & \;\;- & \;\;\tikz\draw[black,fill=black] (0,0) circle (.5ex); & \;\;- & \;\;\tikz\draw[black,fill=black] (0,0) circle (1.ex); & \;\;\tikz\draw[black,fill=black] (0,0) circle (.5ex); & \;\;- & \;\;\tikz\draw[black,fill=black] (0,0) circle (.5ex); & \smalldot  & \smalldot \\
    \bottomrule
\end{tabular}
\caption{Relevance of materials and chemistry simulation areas to climate-related technology. Because different rows tend to require distinct quantum algorithmic approaches, this categorization may guide researchers in choosing areas for developing better algorithms. Note that a particular simulation is likely to be relevant to multiple rows (\textit{e.g.} consider a reaction in solution involving transition metal elements). Dot sizes roughly represent the proportion of application-relevant simulations that would require the given simulation type. Large circles: simulation type very impactful to end-use application; Medium circles: somewhat impactful; hyphen: not impactful.}
\label{fig:simtype_vs_app}
\end{table*}

\subsection{Linking quantum computation to climate applications}
Different physical processes or phenomena require different simulation algorithms. For instance, the approaches used for finding the ground state of an atomic nucleus are different from those for studying dynamics in a solvated chemical process. Algorithmic differences arise partly because the desired quantities or observables are different, but also partly because some aspects of the problem may be tractable on a classical computer, allowing for hybrid computational techniques. It is therefore useful to categorize applications based on the desired physical quantity or observable.

Our main findings are summarized in Table~\ref{fig:simtype_vs_app}, where we display a list of simulation targets and relate them to materials applications in green technology or climate science \cite{Chu16_natmat}. This grid and the information in this document is meant to help guide researchers in developing improved quantum algorithms for particular technologies.

Our approach to Table~\ref{fig:simtype_vs_app} merits some comment, as there are several possible approaches for making connections between quantum computation and end-use technology. It is not immediately obvious which option is most useful for the research community. One could have linked (a) quantum algorithms to application, (b) chemical or material type to application, or (c) a more general ``simulation goal'' to application. Option (a) is not particularly helpful, since most existing Hamiltonian simulation approaches would be used for nearly all applications. Option (b) may be workable but it ignores certain subtleties, like the fact that different properties for the same substance require entirely different algorithmic approaches. The option we chose, option (c), focuses on the simulation ``goal,'' in the sense of both target material \textit{and} desired property. The result is a mix of categorization both by substance (\textit{e.g.} actinides exhibit relativistic effects while most transition metals do not) and by desired property (\textit{e.g.} solution chemistry uses different approaches from gas-phase reactivity).

\subsection{Which climate application will come first?}
It would be useful to determine in detail which climate science applications will be the first to benefit from quantum simulation, although such an analysis remains difficult to do. There are approaches based on the phase-estimation algorithm~\cite{aspuru2005simulated,wecker2014gate,reiher2017elucidating}, which are too costly to be implemented on early generations of quantum hardware. An alternative to this approach is the variational quantum eigensolver (VQE)~\cite{peruzzo2014variational}, which requires the use of a (possibly imperfect) quantum computer and a classical computer working closely together \cite{cerezo12_varalgreview}. Indeed, most ``near-term'' algorithms fall in the category of VQE and are actively being developed by the academic and industrial communities~\cite{preskill2018quantum,bharti2021noisy}.

It is also worth noting that, when designing a quantum algorithm, it is not just the number of qubits but also the \textit{quantum circuit depth} that must be considered. Roughly speaking, the depth equals the number of computational steps that are needed to execute an algorithm. Circuit-depth is currently the main limitation to realizing quantum computation, illustrated by the fact that the best quantum computers cannot reliably achieve a depth greater than 100 steps \cite{arute2019quantum}. Hence new research in this direction must continue to focus on reducing the circuit depth through algorithmic improvements. 

Another rule of thumb is, unsurprisingly, that simpler systems will be simulatable before complex ones. Hence a problem type that is described by an energy function or cost function with a number of terms that grows linearly with system size will likely be successfully treated before a one that grows quadratically, for example. A poor scaling of the number of terms with system size is likely to increase the circuit depth of a simulation. Though it is beyond the scope of this work, we believe the community would benefit from a detailed and accurate categorization of the problem complexities of different materials and different simulation objectives.

It is unlikely that the first useful quantum simulation will be the one that is most relevant to climate, though the possibility that any climate application will benefit from quantum computation is an exciting prospect. One purpose of this document is to guide future researchers in determining which areas to focus attention. 

\newpage
\section{Quantum Optimization \\ \scriptsize{Stuart Hadfield and Michał Stęchły}}
\label{s: QOptimization}

While optimization offers a broad area of potential applications of quantum computing related to climate, at this early stage it remains unclear i) to what degree quantum computers can provide advantages in optimization over classical algorithms in impactful applications of interest, and ii) at what timeframe we can expect to realize such advantages in terms of both hardware and algorithm development. 
Indeed, quantum approaches to optimization remain an active area of research, both in terms of general algorithms and specific applications, for near-term devices and beyond. Moreover, optimization exhibits the potential for different modes of possible quantum advantage, in particular in either settings of exact or approximate optimization (including heuristics), which have different suitability in different applications, as well as sampling problems. While we outline here several potentially impactful areas directly related to the intersection of optimization and climate change, further research is needed to elucidate specific challenging computational tasks in detail, and their implementation on and potential benefit from quantum devices. For example, a recent paper \cite{dalyac_qualifying_2020} considers quantum approaches to problems related to smart-charging of electric vehicles, and 
 observes performance competitive with classical methods. Further work remains to develop the links between such possible computational advantages and the resulting real-world impacts. 
Hence, Q4Climate encourages further interaction between the climate and quantum communities towards leveraging quantum technologies related to optimization. Much research remains to be done, and we are still at the early stages of understanding the power of quantum approaches to optimization, generally. 

\subsection{Quantum hardware and algorithms}
Though recent quantum hardware has begun to allow the preliminary exploration of quantum approaches to optimization, there is a long way to go before these devices can offer broad real-world impact. Two particular challenges are the capability limitations of existing or near-term hardware, and the fact that in many cases we already have very good polynomial time exact or approximate classical algorithms. On the other hand, as these quantum algorithms are believed to not be efficiently classically simulatable in general, the potential for quantum computational advantage remains tantalizing and further research is required to identify the most promising applications. 

The first commercial quantum devices, quantum annealers, were deployed to solve certain classes of hard optimization problems. These devices are severely restricted in the class of algorithms they offer. Despite a decade of experimentation, the scaling advantages offered by such hardware and its future incarnations remains unclear \cite{albash_adiabatic_2018,crosson_prospects_2020}. More recently, quantum gate-model devices have also begun to appear, offering access to a potentially much wider variety of quantum algorithms. We separate, roughly, these algorithms into two classes: 1) near-term quantum algorithms, and  2) fault-tolerant algorithms. Similar to the situation with quantum annealing, near-term algorithms (such as, for example, QAOA~\cite{farhi2014quantum,hadfield_quantum_2019}, or related variational approaches) enable mapping optimization problems to near-term quantum hardware. Despite much recent interest the power, potential advantage, and applicability of these algorithms similarly remains unclear, especially in the context of small noisy real-world devices of the foreseeable future \cite{hadfield_quantum_2019,sanders_compilation_2020}. Hence, we are unable to make concrete claims as to the potential advantages and impact towards climate offered in the near-term by either quantum annealers or gate-model quantum computers. 

Further afield quantum devices, typically considered in a fault-tolerant regime, offer access to a much wider variety of algorithms. All of the previously mentioned near-term algorithms are expected to perform better on fault-tolerant quantum devices. Similarly, better hardware will allow larger problems to be tackled and larger quantum circuits to be reliably deployed (for example, QAOA on $n$ qubits (variables) with number of layers growing as $\log n$ or poly$(n)$). Again, for these methods questions as to what advantages are possible remain promising but unclear. Looking ahead more broadly, the case for quantum advantage is theoretically stronger if we consider algorithms allowed to run for longer than polynomial time in the input size. The standard example is Grover’s algorithm for quantum search which achieves quadratic speedup, and in some cases can obtain a similar speedup of classical algorithms. Recent extensions of these ideas include applications to provable speedups (under certain assumptions) for standard optimization algorithms such as branch-and-bound and backtracking tree search, e.g., in the context of constraint satisfaction problems, mathematical programming, and partial differential equations. As these algorithms have stringent quantum hardware requirements (i.e., sufficiently low error rates and long coherence times), they are much less promising to provide a computational advantage in application to climate change within the near term, though remain an important future research direction. 

In summary, optimization applications related to climate appear a lucrative, though presently somewhat murky, area for potential quantum advantage in the near future. We encourage further research towards identifying the most promising applications, in particular through enhanced dialogue between the quantum and climate communities, but also from industries such as energy, transport, and operations research. We emphasize that as basic research continues in quantum optimization, the viewpoints here could potentially change significantly in the near future to reflect the developing science and technology. In particular, quantum computing research can sometimes lead to new and improved classical algorithms, which is a further welcome development. 

\subsection{Target criteria for optimization applications}

We seek problems and application areas that ideally meet two conditions: i) impactful for helping with climate change, and 
ii) a good fit for quantum resources and potential quantum advantage. 
We identify the following criteria to guide the selection of such problems:
\begin{itemize}
    \item 
    Moderate problem size – the number of problem variables will be hardware-limited for the foreseeable future, as well as the required     algorithm depth and connectivity.
    \item Non-data intensive – applications requiring means of inputting large amounts of classical data (e.g., via qRAM) are less likely to be suitable for near-term approaches or hardware.
    \item Measurable impact – optimization aspects can be one part of large industrial processes, and hence the true  environmental impact of potential improvement might be challenging to measure, and easy to over or understate. Nevertheless in some applications even marginal improvements can have large impact over time or at scale. 
    \item Impact multiplier – how often the problem needs to be solved. For example, quantum computers could be deployed in time-sensitive optimization tasks such as wireless signal decoding~\cite{kim2020towards}, or allocation in distribution networks, to produce better approximate solutions faster. Alternatively, other applications allow a greater computational effort towards obtaining the best solution possible, such as, for example, optimizing the physical layout of a field of wind turbines.
    \item Second-order effects – impact should be gauged within the wider societal and climate contexts, including possible uninteded consquences. (For example, economic effects of improving automobile fuel efficiency could counter-intuitively lead to more cars on the road and an overall increase of emissions...)

\end{itemize}

\subsection{Example applications related to optimization and climate}

We briefly mention five initially promising application areas, including a number of examples from the literature. An immediate goal is to further engage with researchers from the climate science, classical optimization, and quantum computing communities as to better elucidate the potential impact of quantum computers, as described above. 
\begin{itemize}
    \item Power and energy: real-time allocation and routing with immediate potential power and energy savings, e.g., static or dynamic grid optimization problems \cite{li_stochastic_2016,tangpanitanon_use_2019} (including load balancing or unit commitment), batteries~\cite{dalyac_qualifying_2020}, renewable energy \cite{katsigiannis_hybrid_2012,giani2021quantum}, and real-time wireless network~\cite{kim2020towards,choi2020energy}.
    \item System layout: design of facilities, manufacturing plants, and supply chains, e.g., wind turbine placement~\cite{rivas_solving_2009}, hybrid energy system design~\cite{ekren_size_2010,ajagekar2019quantum}, and in 
    automotive industry applications~\cite{luckow2021quantum}. 
    \item Transportation networks: improvements offer direct carbon reductions, e.g., traffic flow and navigation~\cite{nouasse_constraint_2016,neukart2017traffic,imrecke2021maritime}. 
    \item Distribution networks: optimized scheduling and allocation of goods and resources, e.g., potential improved efficiencies in shipping, water distribution \cite{mortazavi-naeini_robust_2015,haguma_water_2015}, buildings and cities \cite{paya-zaforteza_co2-optimization_2009,arunachalam2014quantum}.
    \item Climate mitigation and adaptation: climate \cite{walter_nonlinear_2003} and weather \cite{bardossy_generating_1998} modeling and control \cite{moles_global_2003}, adaption \cite{chuine_fitting_1998}, biodiversity and ecosystems \cite{matear_parameter_1995}, reducing carbon emissions \cite{filar_application_1996}.
\end{itemize}
We remark that an encouraging sign is the deployment of the classical simulated annealing algorithm in applications related to climate, which in some cases may be adapted to give fast quantum algorithms, though further research is required.  Similarly, ongoing developments in applications of classical machine learning and artificial intelligence to climate change \cite{rolnick2019tackling} may lead to or inspire novel use cases for quantum computers. In the other direction, in some cases, quantum optimization research may also lead to novel \lq\lq quantum-inspired\rq\rq\ classical approaches (e.g., \cite{qin2017quantum}). 

\newpage
\section{Quantum Sensing \\ \scriptsize{Tracey Forrest}}
\label{s: QSensing}

Sensors are critical technology in assessing and addressing issues relating to climate change. Current state-of-the-art sensing technologies are already used to monitor several climate related properties, such as atmospheric levels of select greenhouse gas emissions and optical properties of aerosol particles. Quantum sensors bring a new paradigm for measuring their environment, enabling efficiencies and correlations that are not possible in the classical world. Applied to climate change, they are anticipated to have transformative impacts across a variety of domains, from electricity to environmental monitoring.

Sensing covers detection, imaging, metrology and navigation. The properties measured include electric and magnetic fields, motion (including acceleration, rotation and gravity), and optical signals. Through \textit{quantum} sensors, orders of magnitude improvements in sensitivity, selectivity or resource efficiency can potentially be achieved; the power of quantum effectively translates into a dramatic step-function transformation. For example, quantum sensors allow for the measurement of electric and magnetic fields very accurately across many frequencies; drift-free operation, obviating the need for calibration; the ability to sense changes in gravity to reveal potential climate change indicators; non-line-of-sight imaging; and navigation in GPS denied environments.

\subsubsection*{Opportunities for quantum sensing}
The intersection of climate change and quantum sensing represents a vast landscape of potential opportunity. This chapter aims to present an initial scan of applications where quantum may offer an advantage over classical sensors. Applications are identified that assign a high value to one or more of the following:

\begin{itemize}
\item A classical solution does not yet exist. There are very good classical sensors today. Thus, quantum sensors will be driven by applications where they are found to uniquely meet a need.
\item The opportunity calls for greater precision, selectivity or efficiency. This is where the power of quantum may be realized over classical technology.
\item A high tech solution is advisable. Quantum technology is often more complex and costly than classical. A workplace with experience in deploying high tech solutions is more likely to enable successful adoption of quantum technology.
\item Early adopters exist. Demand-side pull provides a focus for technology development, and supports technology validation and commercialization.
\item High impact is expected. Opportunities that offer greater potential to scale, achieve significant emissions reduction and lead to transformational change are of higher priority.
\end{itemize}

Climate change solution domains offer a way of categorizing quantum sensing applications. Five domains have been identified for this purpose as follows: i) electricity systems, ii) transportation, iii) industry, iv) environmental monitoring and v) society.

\subsection{Electricity systems}
Decarbonizing our energy supply is one of the most important ways humanity can achieve sizable reductions in greenhouse gas emissions. Quantum sensors offer a pathway to higher efficiency solar cells and higher solar fuel conversion efficiency through improved materials characterization and quantum coherent approaches, enhanced remote-sensing for the identification of promising geothermal sources \cite{van_der_meer_geologic_2014} or other renewable energy resources \cite{avtar_exploring_2019}, and more effective monitoring of nuclear plants \cite{coble_calibration_2012,crolla_alternative_2009}. Furthermore, while society transitions away from fossil fuels, quantum sensing may help reduce current-system impacts by improving fossil fuel infrastructure maintenance and leak detection.

\subsection{Transportation}
Electric mobility continues to dominate new vehicle investments and is expected to grow significantly in market share. Presently there are technical issues with the diagnosis and prognosis of electric vehicle battery state of health~\cite{li_capacity_2018}, which leads to an opportunity to improve battery performance and lifetime. For example, current state-of-the-art procedures to assess Solid Electrolyte Interphase - a key parameter in battery performance - do not provide an adequate understanding due to measurement difficulties \cite{an_state_2016}. Quantum sensors may help to alleviate these challenges by improving our understanding of battery degradation mechanisms \cite{ford_quantum_nodate}. It’s worth noting that neutron interferometry is already applied to battery diagnostics today.

\subsection{Environmental monitoring}
The monitoring of greenhouse gases is critical to assess the state of climate change \cite{merlone_meteomet_2015,sairanen_validation_2015,verkerke_remote_2014}. Satellites are used in a variety of settings to monitor methane, carbon dioxide and other gases. Accurate readings of methane, in particular, are difficult to obtain because of spectral interference from other gases in spectroscopy, a challenge quantum sensors may overcome. Quantum sensors may also be used to identify and monitor greenhouse gas emissions via satellite over large areas such as peatlands \cite{nations_peatlands_2020}. Moreover, aerosols, which play an important role in climate change, also currently suffer from sensing limitations that may be addressed through quantum technologies \cite{li_characterization_2018}.

Ice sheet melting dynamics and sea-level rise are crucial parts of climate change models. Their monitoring, through optical properties suitable for quantum sensing, would provide important data to climate modelers \cite{kern_antarctic_2016}. In addition, quantum gravimeters deployed via satellite may lead to a better understanding of a broader set of indicators and mechanisms of the global climate system including global mass variations, earth’s response to natural and human-induced forces and monitoring of polar regions for example \cite{tapley_contributions_2019}.

\subsection{Industry}
Monitoring of specific and relevant environmental targets can be used to reduce industrial contributions to climate change. Some promising applications include precision agriculture \cite{maes_estimating_2012} and cattle management \cite{ramayocaldas_identification_2020}. For example, measuring biomarkers of methane emissions in cattle may enable the selection of cattle for breeding which emit less methane and ultimately may lower agricultural greenhouse gas emissions \cite{neethirajan_recent_2017}.

Quantum sensors can be used to image magnetic fields with unprecedented performance, which may in turn lead to an acceleration in the development of smart materials for a variety of applications highly relevant to climate change. For example, today’s expertise in mapping heterogeneous magnetic materials with submicron resolution offers the potential to probe multi-phase magnetic solids for beyond Moore’s law information processing (with commensurate gains to energy performance) \cite{chattopadhyay_quantum_2016}.

\subsection{Society}
Quantum sensors may benefit public health and disaster response efforts. For example, accurate identification of hazards may help inform evacuation planning. Quantum sensors also offer the potential to improve climate risk scenarios by providing new or more accurate climate information on important indicators such as flooding and forest fire prediction.  Quantum gravimeters are already being studied for use in volcano hazard assessments and mitigation plans \cite{carbone_newton-g_2020}. 

The potential applications and benefits of quantum sensing to climate change are broad. Opportunities exist today across electricity, transportation, industry, environmental monitoring, and society, among others. Identifying the most important targets for quantum sensors, and analyzing their enhanced impact for climate change, will serve to guide and catalyze their development.

\newpage
\section{Energy Efficiency of Quantum Computers \\ \scriptsize{Karl Thibault}}
\label{s: EnergyEfficiency}

The potential of using quantum computers to solve problems that are at present impossible on supercomputers is becoming widely known, even in the general public. However, even if quantum computers were not able to achieve a speedup in computing time compared to classical computers for a given problem, using them might still provide an advantage. Indeed, quantum computers could possibly provide an increase in efficiency in terms of the physical resources required, such as the total energy needed to complete a given calculation. Simply comparing two existing devices, the D-Wave 2000Q which consumes 25~kW \bibnote{\url{https://www.dwavesys.com/sites/default/files/D-Wave\%202000Q\%20Tech\%20Collateral_0117F_0.pdf}} and the Summit supercomputer which consumes 13~MW of power \cite{noauthor_summit_2021}, shows a difference in energy consumption of approximately four orders of magnitude. This lower energy cost comes from lower cooling requirements, as well as lower power needed to operate the computing units. Saving megawatt-hours of energy drastically lowers the environmental footprint of computing, as well as saving millions of dollars.

However, it is important to note that the D-Wave device and a supercomputer differ significantly in the type of problems they can solve. We classify quantum computing devices in three categories: single-purpose devices, Noisy Intermediate-Scale Quantum (NISQ) devices \cite{preskill_quantum_2018} and fault-tolerant devices. \textit{Single-purpose devices}, such as the D-Wave 2000Q, are designed to implement a specific algorithm, in this case quantum annealing. \textit{NISQ} devices are made up of, as their name implies, a limited amount of imperfect qubits over which we exercise imperfect control. Finally, a \textit{fault-tolerant quantum computer's}  qubits are protected from errors using quantum error correction and an appropriate hardware design. Fault-tolerant devices are the most directly comparable to today's supercomputers, but unfortunately remain far from attainable in the near term.

Indeed, because we are still in the NISQ era, comparisons such as the one made above are often considered murky. However, these are the only comparisons that can currently be made accurately, rather than speculatively. We will thus focus our attention on comparing classical and quantum devices on a specific demonstrated computation that has already been achieved in the literature. For this purpose, we have chosen the quantum supremacy experiment, where Google claimed that their quantum computer provided a speedup over Summit \cite{arute_quantum_2019}. This report aims to assess the efficiency difference between classical and quantum computation in this specific example. However, comparing the efficiency of two different devices is a complex task. To simplify it, the lifetime of a specific device can be split into three parts: production, operation and end-of-life. Here, we will focus on the \textit{operation} of the devices. Some brief comments will be made regarding production and end-of-life considerations.

\subsection{Operating costs}
Operating costs can be separated into four categories: electricity, facilities, personnel and hardware/software.

Electricity costs (including cooling) are calculated as follows:
\begin{center}
    Operating time * power use * cost of kWh = electricity costs.
\end{center}
Let's assume 0.1~\$/kWh for electricity costs. Let's also assume that Google's quantum supremacy calculation can be done in 200~seconds on a quantum computer \cite{arute_quantum_2019} and in 2.5~days on Summit \cite{pednault_quantum_2019}. Finally, let's assume 25~kW of power usage for the quantum computer, which is a good approximation if it is a superconducting chip housed inside a dilution refrigeration (taking into account cooling and all operation electronics)~\bibnote{A dilution refrigeration's most power hungry component is the compressor, which typically consumes less than 15~kW. Adding control devices for the cryostat itself, its power consumption can go up to $\sim20$~kW (or more if a magnet is in use). Finally, taking into account the electronic devices to control the chip inside the cryostat, the total power consumption is normally less than 25~kW. This is in line with the D-Wave's 2000Q's power consumption (see ref. [90])}. This is to be compared with Summit's 13~MW of power usage~\cite{noauthor_summit_2021}.

\begin{center}
    For the classical computer, we obtain: \, 60 h * 13 000 kW * 0.1 \$/kWh = 78,000 \$.\\ 
    \vskip 0.5pc
    For the quantum computer, we obtain: \hskip 2pc 0.056 h * 25 kW * 0.1 \$/kWh = 0.14 \$.
\end{center}

Quantum computation of this specific problem costs 557~000~times less than the same computation done on a supercomputer. Observe that this is in part due to the lower power consumption of quantum computers, not strictly because the quantum computation is faster, as if both computations took the same time the quantum computation would still cost 520~times less. Without specifying the problem to be solved and the speed of computation, operational costs for a full year are 21.9~k\$ for a superconducting quantum computer vs 11.39~M\$ for Summit. Even if Google's quantum supremacy's claim was incorrect and their quantum computer did not provide a speedup, the fact that it could compete with Summit, the world's fastest supercomputer at the time, is an incredible feat in itself. Considering that Google's chip fits into a dilution refrigerator and needs less than one hundred laboratory grade electronic machines to operate --- compared to Summit's 9216~CPUs, 27648~GPUs and other classical control components --- it definitely seems more power efficient. Moreover, power consumption should not scale rapidly with the number of qubits, at least initially, as a chip with some thousands of qubits can still fit in a dilution fridge.

Facilities costs include rent and water usage costs. They are similar for both types of computers, although we can expect that quantum computers will need much less space than supercomputers do as of today. Even if some predictions of fault-tolerant computers anticipate that their initial size could be as huge as many football fields \cite{lekitsch_blueprint_2017}, one can hope that further development will lead to decreased size just as for classical computers in the past decades. This report will not dive deeper into this area, as it is yet unclear if there is a gain to achieve in this area. If any, this gain will surely be much less than for electricity costs. One specific area which might be worth studying in more details, because of its environmental impact, is the difference in water consumption for cooling and other purposes such as fabrication. This is however out of the scope of this report.

Personnel costs are similar for both types of computers, as both need a team of ten to fifty people to run. This amounts to $\sim$1-5~M\$ per year. Gains are not expected in this area, nor does it have much of an impact on the environment per se.

Hardware and software costs are mostly for buying the equipment necessary to run the computers, as well as maintenance costs to repair dysfunctional parts. First of all, we suppose software costs to be minimal compared to the hardware, as the operation of software is in essence free once acquired (or subject to typically negligible recurring fees).  On the hardware side, maintenance costs can be substantial, but are negligible compared to the upfront costs over a reasonable length of time. Indeed, Summit for example is composed of 27648~GPUs and 9216~CPUs, for which the total cost amounts to $\sim325$~M\$ \cite{noauthor_summit_2021}. Quantum computers, on the other hand, cannot yet simply be bought. One can estimate the costs of all the cooling ($\sim1$~M\$) and electronics to operate them (10-20~M\$), but the cost of the actual chip is yet undetermined. It is thus currently impossible to assess the cost difference accurately. 

Finally, this assessment does not take into account research and development of new hardware or software, which are substantial investments for both types of computers.

\subsection{Other factors}
This assessment covers only a small portion of the footprint of quantum computers. In particular, it highlights the reduction in electricity consumption that could be achieved from operating quantum computers versus classical supercomputers. More thorough studies are needed to clarify the difference in overall operating costs, but also in fabrication and environmental costs such as non-renewable material or heavy metal use, as well as the biodiversity and ecosystem impact of the fabrication chain including the end-of-life phase. Even if chips themselves need approximately the same amount of heavy or non-renewable materials to create, quantum computers of today have only one chip compared to the tens of thousands of chips in supercomputers. This could be a huge reduction in the use of those materials but, in the long-term, quantum error correction schemes might need a very large number of classical and/or quantum chips to operate, counterbalancing this reduction. In any case, there is a very clear need for deeper analysis of the advantage of using quantum devices, not for their speed, but their efficiency.

\newpage
\section{Conclusion}
\label{s:conclusion}
As quantum technologies continue to develop, it is important to assess how they may be best utilized to provide real-world impact. Here, we argue that the advent of large and accurate quantum computers could provide efficient methods to simulate a host of challenging physical systems that are crucial to tackle climate-change related problems in energy, industrial processes, atmospheric science, and other sectors. Similarly, quantum approaches to optimization could provide benefits in areas including renewable energy, systems layout, transportation, distribution, and direct climate modeling and mitigation, though their precise advantages over classical computers requires further investigation. More generally, it could be useful to use quantum computers not only for their speed-up in certain use-cases, but for their improved energy efficiency. Quantum sensors could be relevant for climate mitigation, as they can provide improved sensitivity, selectivity and efficiency gains across a wide variety of application domains. 
In each case, it is important to continue to identify and analyze specific applications having the greatest possible impact, as well as investigate technological innovations such as improved algorithms and hardware towards enabling their successful implementation as soon as possible. 

Promising areas not discussed in this preliminary assessment include quantum machine learning and quantum approaches for solving differential equations, among others.
We are optimistic that increased dialogue between the quantum and climate science communities will help to further identify useful quantum approaches for climate, as well as critical bottlenecks where quantum technologies could be the most effective.

The scope and alarming pace of climate change only serves to underscore the importance and urgency of drawing the promise of advanced technologies closer to climate impact today. 
Indeed, the ongoing development of quantum technologies is contemporaneous with the adverse effects of a changing climate; we are hopeful the former wins the race. 


\section*{Acknowledgements}
We acknowledge the guidance and contribution of Q4Climate's advisory board members: Alán Aspuru-Guzik, Alexandre Blais, Ron Dembo, Nicolas Farina, Michele Mosca, Kristin Persson, Brian Chol Soo Standen and Krysta Svore. We thank the members of the ClimateChangeAI initiative David Rolnick and Priya L. Donti. We also thank  Monika Barcikowska, Alexandru Petrescu, Jean-Pierre Blanchet and Christian Sarra-Bournet for fruitful discussions.

\bibliographystyle{unsrt}
\bibliography{references}
\appendix

\end{document}